\RequirePackage{lineno}
\documentclass[aps,prl,twocolumn,superscriptaddress,groupedaddress,preprintnumbers]{revtex4}  
\usepackage{graphicx}  
\usepackage{dcolumn}   
\usepackage{bm}        
\usepackage{amssymb}   
\usepackage{multirow}
\usepackage{epsfig}
\usepackage{amsmath}
\usepackage[colorlinks, linkcolor=blue]{hyperref}
\usepackage{ulem}
\usepackage{appendix}

\newcommand{\effstw}{\ensuremath{\sin^2\theta_{\text{eff}}^{\text{$\ell$}}}}

\begin{document}
\lefthyphenmin=2
\righthyphenmin=3

\widetext
\hspace{5.2in} \mbox{FERMILAB-PUB-24-0106}

\title{Up and Down Quark Structure of the Proton}
%
\affiliation{LAFEX, Centro Brasileiro de Pesquisas F\'{i}sicas, Rio de Janeiro, RJ 22290, Brazil}
\affiliation{Universidade do Estado do Rio de Janeiro, Rio de Janeiro, RJ 20550, Brazil}
\affiliation{Universidade Federal do ABC, Santo Andr\'e, SP 09210, Brazil}
\affiliation{University of Science and Technology of China, Hefei 230026, People's Republic of China}
\affiliation{Universidad de los Andes, Bogot\'a, 111711, Colombia}
\affiliation{Charles University, Faculty of Mathematics and Physics, Center for Particle Physics, 116 36 Prague 1, Czech Republic}
\affiliation{Czech Technical University in Prague, 116 36 Prague 6, Czech Republic}
\affiliation{Institute of Physics, Academy of Sciences of the Czech Republic, 182 21 Prague, Czech Republic}
\affiliation{Universidad San Francisco de Quito, Quito 170157, Ecuador}
\affiliation{LPC, Universit\'e Blaise Pascal, CNRS/IN2P3, Clermont, F-63178 Aubi\`ere Cedex, France}
\affiliation{LPSC, Universit\'e Joseph Fourier Grenoble 1, CNRS/IN2P3, Institut National Polytechnique de Grenoble, F-38026 Grenoble Cedex, France}
\affiliation{CPPM, Aix-Marseille Universit\'e, CNRS/IN2P3, F-13288 Marseille Cedex 09, France}
\affiliation{LAL, Univ. Paris-Sud, CNRS/IN2P3, Universit\'e Paris-Saclay, F-91898 Orsay Cedex, France}
\affiliation{LPNHE, Universit\'es Paris VI and VII, CNRS/IN2P3, F-75005 Paris, France}
\affiliation{IRFU, CEA, Universit\'e Paris-Saclay, F-91191 Gif-Sur-Yvette, France}
\affiliation{IPHC, Universit\'e de Strasbourg, CNRS/IN2P3, F-67037 Strasbourg, France}
\affiliation{IPNL, Universit\'e Lyon 1, CNRS/IN2P3, F-69622 Villeurbanne Cedex, France and Universit\'e de Lyon, F-69361 Lyon CEDEX 07, France}
\affiliation{III. Physikalisches Institut A, RWTH Aachen University, 52056 Aachen, Germany}
\affiliation{Physikalisches Institut, Universit\"at Freiburg, 79085 Freiburg, Germany}
\affiliation{II. Physikalisches Institut, Georg-August-Universit\"at G\"ottingen, 37073 G\"ottingen, Germany}
\affiliation{Institut f\"ur Physik, Universit\"at Mainz, 55099 Mainz, Germany}
\affiliation{Ludwig-Maximilians-Universit\"at M\"unchen, 80539 M\"unchen, Germany}
\affiliation{Panjab University, Chandigarh 160014, India}
\affiliation{Delhi University, Delhi-110 007, India}
\affiliation{Tata Institute of Fundamental Research, Mumbai-400 005, India}
\affiliation{University College Dublin, Dublin 4, Ireland}
\affiliation{Korea Detector Laboratory, Korea University, Seoul, 02841, Korea}
\affiliation{CINVESTAV, Mexico City 07360, Mexico}
\affiliation{Nikhef, Science Park, 1098 XG Amsterdam, the Netherlands}
\affiliation{Radboud University Nijmegen, 6525 AJ Nijmegen, the Netherlands}
\affiliation{Joint Institute for Nuclear Research, Dubna 141980, Russia}
\affiliation{Institute for Theoretical and Experimental Physics, Moscow 117259, Russia}
\affiliation{Moscow State University, Moscow 119991, Russia}
\affiliation{Institute for High Energy Physics, Protvino, Moscow region 142281, Russia}
\affiliation{Petersburg Nuclear Physics Institute, St. Petersburg 188300, Russia}
\affiliation{Instituci\'{o} Catalana de Recerca i Estudis Avan\c{c}ats (ICREA) and Institut de F\'{i}sica d'Altes Energies (IFAE), 08193 Bellaterra (Barcelona), Spain}
\affiliation{Uppsala University, 751 05 Uppsala, Sweden}
\affiliation{Lancaster University, Lancaster LA1 4YB, United Kingdom}
\affiliation{Imperial College London, London SW7 2AZ, United Kingdom}
\affiliation{The University of Manchester, Manchester M13 9PL, United Kingdom}
\affiliation{University of Arizona, Tucson, Arizona 85721, USA}
\affiliation{University of California Riverside, Riverside, California 92521, USA}
\affiliation{Florida State University, Tallahassee, Florida 32306, USA}
\affiliation{Fermi National Accelerator Laboratory, Batavia, Illinois 60510, USA}
\affiliation{University of Illinois at Chicago, Chicago, Illinois 60607, USA}
\affiliation{Northern Illinois University, DeKalb, Illinois 60115, USA}
\affiliation{Northwestern University, Evanston, Illinois 60208, USA}
\affiliation{Indiana University, Bloomington, Indiana 47405, USA}
\affiliation{Purdue University Calumet, Hammond, Indiana 46323, USA}
\affiliation{University of Notre Dame, Notre Dame, Indiana 46556, USA}
\affiliation{Iowa State University, Ames, Iowa 50011, USA}
\affiliation{University of Kansas, Lawrence, Kansas 66045, USA}
\affiliation{Louisiana Tech University, Ruston, Louisiana 71272, USA}
\affiliation{Northeastern University, Boston, Massachusetts 02115, USA}
\affiliation{University of Michigan, Ann Arbor, Michigan 48109, USA}
\affiliation{Michigan State University, East Lansing, Michigan 48824, USA}
\affiliation{University of Mississippi, University, Mississippi 38677, USA}
\affiliation{University of Nebraska, Lincoln, Nebraska 68588, USA}
\affiliation{Rutgers University, Piscataway, New Jersey 08855, USA}
\affiliation{Princeton University, Princeton, New Jersey 08544, USA}
\affiliation{State University of New York, Buffalo, New York 14260, USA}
\affiliation{University of Rochester, Rochester, New York 14627, USA}
\affiliation{State University of New York, Stony Brook, New York 11794, USA}
\affiliation{Brookhaven National Laboratory, Upton, New York 11973, USA}
\affiliation{Langston University, Langston, Oklahoma 73050, USA}
\affiliation{University of Oklahoma, Norman, Oklahoma 73019, USA}
\affiliation{Oklahoma State University, Stillwater, Oklahoma 74078, USA}
\affiliation{Oregon State University, Corvallis, Oregon 97331, USA}
\affiliation{Brown University, Providence, Rhode Island 02912, USA}
\affiliation{University of Texas, Arlington, Texas 76019, USA}
\affiliation{Southern Methodist University, Dallas, Texas 75275, USA}
\affiliation{Rice University, Houston, Texas 77005, USA}
\affiliation{University of Virginia, Charlottesville, Virginia 22904, USA}
\affiliation{University of Washington, Seattle, Washington 98195, USA}
\author{V.M.~Abazov} \affiliation{Joint Institute for Nuclear Research, Dubna 141980, Russia}
\author{B.~Abbott} \affiliation{University of Oklahoma, Norman, Oklahoma 73019, USA}
\author{B.S.~Acharya} \affiliation{Tata Institute of Fundamental Research, Mumbai-400 005, India}
\author{M.~Adams} \affiliation{University of Illinois at Chicago, Chicago, Illinois 60607, USA}
\author{T.~Adams} \affiliation{Florida State University, Tallahassee, Florida 32306, USA}
\author{J.P.~Agnew} \affiliation{The University of Manchester, Manchester M13 9PL, United Kingdom}
\author{G.D.~Alexeev} \affiliation{Joint Institute for Nuclear Research, Dubna 141980, Russia}
\author{G.~Alkhazov} \affiliation{Petersburg Nuclear Physics Institute, St. Petersburg 188300, Russia}
\author{A.~Alton$^{a}$} \affiliation{University of Michigan, Ann Arbor, Michigan 48109, USA}
\author{A.~Askew} \affiliation{Florida State University, Tallahassee, Florida 32306, USA}
% ...
%
%
\collaboration{The D0 Collaboration\footnote{with visitors from
$^{a}$Augustana University, Sioux Falls, SD 57197, USA,
$^{b}$The University of Liverpool, Liverpool L69 3BX, UK,
$^{c}$Deutshes Elektronen-Synchrotron (DESY), Notkestrasse 85, Germany,
$^{d}$CONACyT, M-03940 Mexico City, Mexico,
$^{e}$SLAC, Menlo Park, CA 94025, USA,
$^{f}$University College London, London WC1E 6BT, UK,
$^{g}$Centro de Investigacion en Computacion - IPN, CP 07738 Mexico City, Mexico,
$^{h}$Universidade Estadual Paulista, S\~ao Paulo, SP 01140, Brazil,
$^{i}$Karlsruher Institut f\"ur Technologie (KIT) - Steinbuch Centre for Computing (SCC),
D-76128 Karlsruhe, Germany,
$^{j}$Office of Science, U.S. Department of Energy, Washington, D.C. 20585, USA,
$^{m}$University of Maryland, College Park, MD 20742, USA,
$^{n}$European Orgnaization for Nuclear Research (CERN), CH-1211 Geneva, Switzerland,
$^{o}$Purdue University, West Lafayette, IN 47907, USA,
$^{p}$Institute of Physics, Belgrade, Belgrade, Serbia,
and
$^{q}$P.N. Lebedev Physical Institute of the Russian Academy of Sciences, 119991, Moscow, Russia.
$^{\ddag}$Deceased.
}} \noaffiliation
\vskip 0.25cm

\begin{abstract}
We measure proton structure parameters sensitive primarily to valence quarks using 8.6 fb$^{-1}$ of data 
collected by the D0 detector in $\sqrt{s}=1.96$ TeV $p\bar{p}$ collisions at the Fermilab Tevatron. 
We exploit the property of the forward-backward asymmetry in dilepton events to be 
factorized into distinct structure parameters and electroweak
quark-level asymmetries. Contributions to the asymmetry from $s$, $c$ and $b$ quarks, as well as from $u$ and $d$ 
sea quarks, are suppressed allowing valence $u$ and $d$ quarks to be separately determined. We find an $u$ to $d$ 
quark ratio near the peak values 
in the quark density distributions that is smaller than predictions from modern parton distribution 
functions.

\end{abstract}
\maketitle

The forward-backward asymmetry, $A_{FB}$, in dilepton production at 
hadron colliders is due to parity violation in the electroweak interaction but also 
depends upon the hadron's partonic structure~\cite{Early1, Early2, Early3, AFBCorrelation}. 
Although many observables depend upon the experimentally indistinguishable contributions from 
different quark flavors, $A_{FB}$ has the capability to provide information on specific quarks. 
Contributions to $A_{FB}$ from the $s$, $c$ and $b$ quarks are significantly suppressed because 
the quark and antiquark densities are nearly the same and thus $A_{FB}$  
predominately depends on $u$ and $d$ quark densities. Moreover, 
the asymmetries for $u\bar{u}$ and $d\bar{d}$ initial states depend differently on the 
dilepton mass ($M$), offering the possibility to obtain $u$ and $d$ quark densities individually. 
Recent analyses~\cite{AFBFactorization, AFBSlope} show that $A_{FB}$ can be factorized 
into separate electroweak and quantum chromodynamics (QCD) functions allowing independent 
determinations of the effective weak mixing angle parameter $\sin^2\theta^{\ell}_\text{eff}$, and proton structure 
parameters called $P_u$ and $P_d$ for $u$ and $d$ quarks, respectively. 
Measurements of $P_u$ and $P_d$ provide unique information about 
the proton structure. 

In this paper we report a determination of $P_u$ and $P_d$ from the $A_{FB}$ distributions in  
$p\bar{p}\rightarrow Z/\gamma^* \rightarrow \ell^+\ell^-$ events using data corresponding to 8.6 fb$^{-1}$ 
of integrated luminosity collected with the D0 
detector at the Fermilab Tevatron $p\bar{p}$ collider at $\sqrt{s}=1.96$ TeV. 
A previous analysis~\cite{CMSD0Extraction} extracted $P_u$ and $P_d$ from the $A_{FB}$ distributions 
measured by the D0 collaboration using 5 fb$^{-1}$ of data in only the dielectron final state~\cite{D05fbAFB} 
and after unfolding the measured mass dependence of $A_{FB}$ to the parton level. 
It demonstrated the feasibility of such a measurement and showed a tendency for $P_d$ to be 
higher and $P_u$ to be lower than expected~\cite{CMSD0Extraction}. 
In this paper a larger data sample 
and both dielectron and dimuon final states are used, 
thus improving the statistical precision relative to Ref.~\cite{CMSD0Extraction}. 
The dilepton mass distributions are not unfolded, thus removing a significant source of systematic 
uncertainty. 
As explained 
below, $P_u$ and $P_d$ in $p\bar{p}$ collisions are dominated by the valence $u$ and 
$d$ quark contributions, and their ratio, $R=P_u/P_d$, directly reflects the relative contributions 
of the two leading quarks inside 
a proton. 
$P_u$, $P_d$ and $R$ can also be measured using the data collected at the LHC, 
but measurements in $pp$ collisions also involve sea quark contributions comparable to 
those of the valence 
quarks. 
The measurement presented in this paper is thus unique and 
provides novel information 
on the valence $u$ and $d$ quark distributions by separating them from each other and suppressing 
heavy quark contributions. 
In addition, since the sum rules in the global analysis of the parton distribution functions (PDFs) 
relate the valence and sea quarks, 
this measurement could have implications for the PDFs of sea quarks and testing the calculations 
of initial state gluon radiation.

At the Tevatron $A_{FB}$ is defined as:
\begin{eqnarray}
 A_{FB} = \frac{N_F - N_B}{N_F + N_B}, 
\end{eqnarray}
where $N_F$ and $N_B$ are the number of forward and backward events,  
defined as those for which $\cos\theta>0$ and $\cos\theta<0$, 
with $\theta$ defined as the angle 
between the direction of the negatively charged lepton and the direction of the proton beam in the 
Collins-Soper frame~\cite{cs-frame}. 
At specific values of the 
dilepton rapidity $Y$ and transverse momentum $Q_T$ defined with respect to the 
beam axis,  
the observed $A_{FB}$ distribution as a function of the 
dilepton invariant mass $M$ can be 
factorized as~\cite{AFBFactorization}:

\begin{footnotesize}
\begin{eqnarray}\label{eq:AFBfactorization}
 A_{FB}(M) &=& 
     \frac{\sum\limits_{q=u,c} [1-2D_q(M) ]\sigma_q(M)}{\sigma_\text{total}(M)} \cdot A^u_{FB}(M) \nonumber \\
 & & + \frac{\sum\limits_{q=d,s,b}[1-2D_q(M)]\sigma_q(M)}{\sigma_\text{total}(M)} \cdot A^d_{FB}(M) \nonumber \\
 &\equiv& \mathcal{C}_u(M)A^u_{FB}(M) + \mathcal{C}_d(M)A^d_{FB}(M), 
\end{eqnarray}
\end{footnotesize}

\noindent where $\sigma_q$ is the 
subprocess cross section for a specific $q\bar{q}$ $(q=u,d,s,c,b)$ initial state,  
$\sigma_\text{total}$ is the total cross section $\sum_{q=u,d,s,c,b}\sigma_q$, and 
$A^u_{FB}$ and $A^d_{FB}$ are asymmetries for initial up-type states ($u\bar{u}$ and $c\bar{c}$) 
and down-type states ($d\bar{d}$, $s\bar{s}$ and $b\bar{b}$), respectively.
Forward and backward events for the $q\bar{q}$ subprocesses 
are defined in the Collins-Soper frame 
in terms of a new angle $\theta'$ between the negatively charged lepton direction 
and the quark direction.
$A^u_{FB}$ and $A^d_{FB}$ are determined by $\sin^2\theta^\ell_\text{eff}$ and are independent 
of parton densities. The dilution factor $D_q$ is defined as the probability for the $q\bar{q}$ subprocess 
to have an initial state where $q$ comes from the antiproton 
while $\bar{q}$ comes from the proton, for which $\cos\theta = -\cos\theta'$. 
The weights for the up- and down-type quarks, $\mathcal{C}_u$ and $\mathcal{C}_d$,  
can be averaged over a finite mass range, to further 
separate them into mass-averaged structure parameters ($P_u$ and $P_d$) and mass-dependent 
structure parameters ($\Delta_u$ and $\Delta_d$)~\cite{AFBFactorization}:
\begin{eqnarray}\label{eq:averagedef}
 \mathcal{C}_{u,d}(M) &=& P_{u,d} + \Delta_{u,d}(M).
\end{eqnarray}
In this Letter, we have defined $P_u$ and $P_d$ by averaging over the mass range of [70, 116] GeV.  
The structure parameters, cross sections, asymmetries, and dilution factors all
 depend on $Y$ and $Q_T$. Note that Eq.~\eqref{eq:AFBfactorization} factorizes the 
 QCD part of the observed $A_{FB}$ into $\mathcal{C}_u$ and $\mathcal{C}_d$, and  
 the electroweak part as $A^u_{FB}$ and $A^d_{FB}$.
 
 The dilution factors $D_u$ and $D_d$ are modeled by the PDFs and 
 are small since the interactions of an antiquark in the proton and a quark in the antiproton
 are suppressed in the relevant $x$-range at the Tevatron. The dilution factors for $s$, $c$ and $b$ quarks are very close 
 to 0.5~\cite{CT18NNLO, MSHT20, NNPDF4} and thus $P_u$ and $P_d$ are dominated by the valence $u$ and $d$ quarks at leading order. 
 As a result, $P_u$ and $P_d$ at the Tevatron are approximately
 \begin{eqnarray}\label{eq:PuPdapprox}
  P_u &\sim& u(x_1)u(x_2) / \sigma_\text{total}(x_1,x_2), \nonumber \\
  P_d &\sim& d(x_1)d(x_2) / \sigma_\text{total}(x_1,x_2), 
\end{eqnarray}
where $x_{1,2}$ is the Bjorken variable for the 
colliding quark and antiquark respectively, defined at leading order as 
$x_{1,2} = \frac{\sqrt{M^2 + Q^2_T}}{\sqrt{s}} e^{\pm Y}$. 
The ratio $R=P_u/P_d$, in which the total cross section cancels, represents the 
relative contribution of $u$ and $d$ quarks. 
Due to the detector acceptance discussed below, the data in this measurement has 
dilepton rapidity in the interval $|Y|=[0, 2.3]$. The 
$P_u$, $P_d$ and $R$ measured in this paper correspond to the values of $x$ from approximately  
0.004 to 0.45. We obtain information on the $x$-dependence of the structure parameters 
by analyzing the data separately for $|Y|$ intervals of [0, 0.5], [0.5, 1.0], [1.0, 1.5], and [1.5, 2.3].

\begin{figure}[!hbt]
\begin{center}
\epsfig{scale=0.43, file=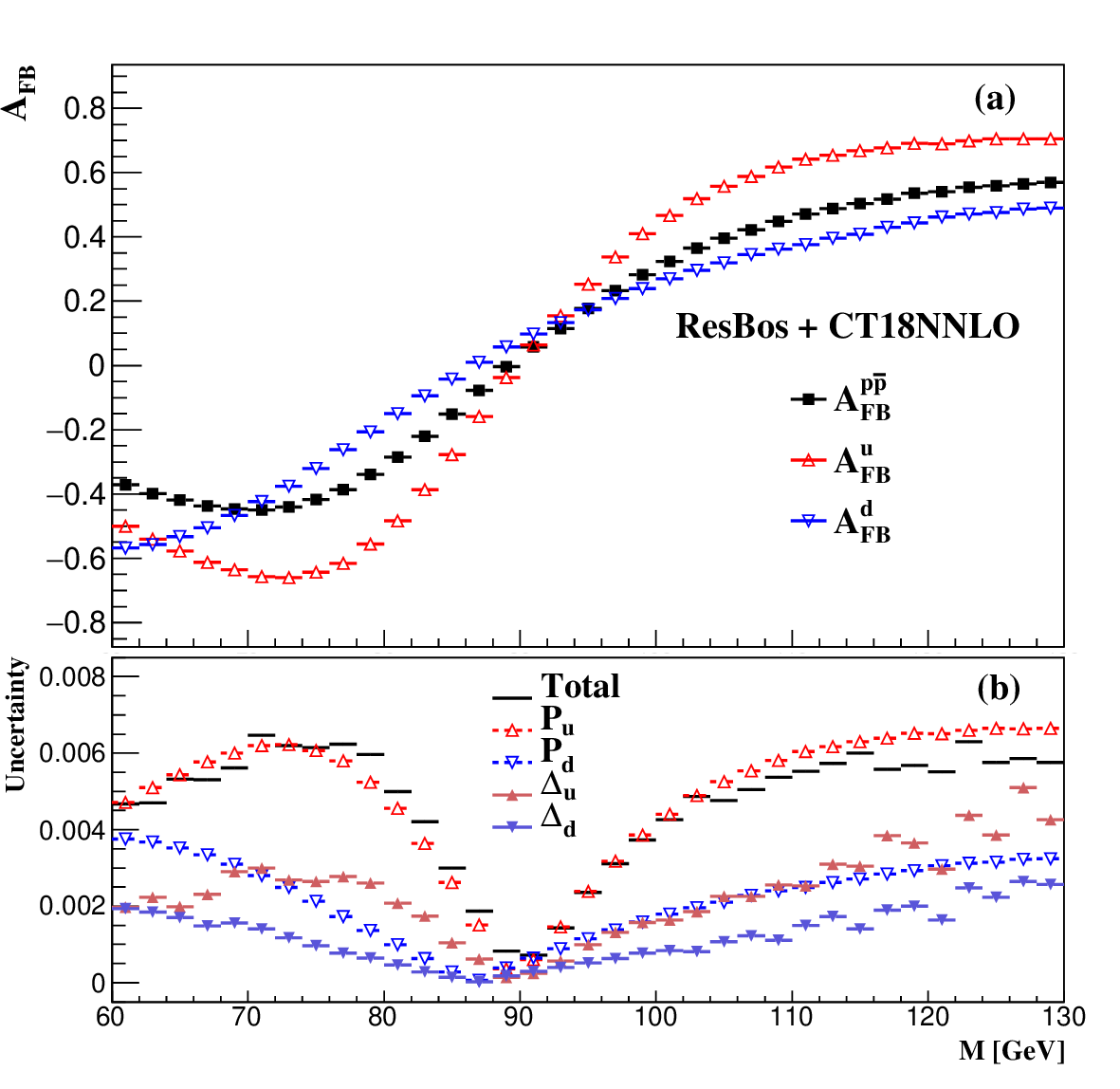}
\caption{\footnotesize (a) The PDF-independent $A^u_{FB}$ and $A^d_{FB}$ predicted by 
{\sc ResBos} as a function of $M$ and the resulting $A_{FB}$ in $p\bar{p}$ collisions using the CT18NNLO PDF. 
(b) The PDF induced absolute uncertainties in $A_{FB}$ due to $P_u$, $P_d$, $\Delta_u$ and $\Delta_d$.}
\label{fig:AFB}
\end{center}
\end{figure}

This Letter focuses on the measurement of $P_u$ and $P_d$. The $\Delta_u$ and $\Delta_d$ 
terms can be predicted with small uncertainties 
for $M$ in a narrow window around the $Z$ boson pole~\cite{AFBFactorization, AFBSlope}. 
$A^u_{FB}$ and $A^d_{FB}$ 
can be precisely predicted and have different dependences on $M$. 
$P_u$ and $P_d$ 
can be determined by comparing Eq.~\eqref{eq:AFBfactorization} to the measured $A_{FB}$ distribution. The asymmetries $A_{FB}^{u}$ and $A_{FB}^{d}$, and the uncertainties on $A_{FB}$ due
to the $P_{q}$ and $\Delta_q$ parameters are 
calculated using {\sc ResBos}~\cite{resbos} with CT18NNLO~\cite{CT18NNLO} PDFs, 
and are shown in Fig.~\ref{fig:AFB}.

The D0 detector consists of a tracking system surrounded by a solenoid magnet, calorimeters, and a muon 
system~\cite{D0det1, D0det2, D0det3}. 
 Dielectron and dimuon events are collected with lepton triggers and are required to have 
a lepton-antilepton pair in the offline analysis.
Leptons are required to be well separated from other particles both in the tracking system and the calorimeter. 
Muons are measured as tracks in the tracking and muon systems with $|\eta_\text{det}|<1.8$~\cite{D0eta}, 
and are required to have transverse momentum $p_T>15$ GeV. 
Electrons are reconstructed as clusters in the 
central calorimeter (CC) 
with $|\eta_\text{det}|<1.1$, and in an end calorimeter (EC) 
with $1.5<|\eta_\text{det}|<3.5$. 
They are required to have a spatially matched track in the tracking system, so that 
their electric charge can be determined, and also for 
discriminating against photons. 
The EC-EC events, where both electrons are in an EC, are 
excluded due to the high level of background for such events. 
The threshold for the electron transverse momentum is 25 GeV. 
As a result, the background contributions  
from $Z/\gamma^*\rightarrow \tau\tau$, 
$W+$jets, diboson ($WW$ and $WZ$), $\gamma\gamma$, top quarks and multi-jets are suppressed to 
$\mathcal{O}(1\%)$ in the mass region 70 $<$ $M$ $<$ 116 GeV 
used in this analysis. 

A Monte Carlo (MC) sample of $Z/\gamma^*\rightarrow \ell^+\ell^-$  
events is generated using the leading-order {\sc pythia} generator~\cite{pythia} 
with CT18NNLO PDFs, followed by a 
{\sc geant}-based~\cite{GEANT} simulation of the D0 detector. The samples are further corrected 
by reweighting the MC events at the generator level in $M$, $Q_T$, $Y$ and 
$\cos\theta$ to match the calculation of {\sc ResBos}~\cite{resbos}, which is at 
approximate next-to-next-to-leading order and next-to-next-to-leading logarithm in QCD. 
The electron energy and muon momentum are calibrated 
using the known resonances in the 
dilepton mass spectrum. The efficiencies of the online and offline selection criteria are determined using 
the tag-and-probe method~\cite{tag-and-probe} and the MC simulation is corrected to be consistent with the data.
The multi-jets background is estimated 
using data, while other backgrounds are determined using 
{\sc pythia} MC simulations.
The methodologies used to derive the energy and 
momentum calibrations, efficiencies and estimations of the background contributions 
were also employed in the previous measurements of the effective weak 
mixing angle~\cite{D0stwee, D0stwmu}. 
Many systematic effects are  
suppressed since $A_{FB}$ is defined as a ratio. 

For the measurement of $P_u$ and $P_d$ in the full $0<|Y|<2.3$ range or in a 
particular $|Y|$ interval, a set of MC template distributions of $A_{FB}$ is prepared in 
which $P_u$ and $P_d$ are varied while keeping $\Delta_u$ and 
$\Delta_d$ fixed at their values calculated using {\sc ResBos} and CT18NNLO. 
A set of $\mathcal{C}_q=P_q + \Delta_q$ values is calculated for intervals 
in $Y$, $M$ and $Q_T$~\cite{Binning}. $A_{FB}$ templates are acquired by 
reweighting the generator level differential cross sections $\sigma_q(Y, M, Q_T, \cos\theta)$ of 
the MC sample according to the $\mathcal{C}_q$ value. 
In the MC reweighting procedures, 
$A^u_{FB}$ and $A^d_{FB}$ are calculated using {\sc ResBos}, with $\effstw$ set to the average of 
the results from the electron-positron colliders LEP and SLC~\cite{LEP-SLD}. Corresponding uncertainties on 
$\effstw$ are extrapolated to the measured 
$P_u$ and $P_d$. We do not use the hadron collider results on $\effstw$ 
in order to avoid the influence from the specific PDF predictions used in their measurement, 
but this choice has a negligible impact on the result because 
the hadron collider measurements~\cite{TevatronStw, ATLASstw, CMSstw, LHCbstw} 
give values of $\effstw$ very close to the combined LEP/SLC result. 
Uncertainties on $\Delta_u$ and $\Delta_d$ are estimated using the 
error PDF sets given by CT18NNLO.   
Equation~\eqref{eq:AFBfactorization} is only strictly true when $Y$ and $Q_T$ dependences are fully considered. 
In this letter, the observed $A_{FB}$ is averaged over $Q_T$ and $Y$ so that the factorization formalism of 
Eq.~\eqref{eq:AFBfactorization} becomes an approximation. This gives rise to additional 
uncertainties in the calculation of $\sigma_q$ and higher order QCD contributions. Part of this uncertainty is 
already included when taking the CT18NNLO error PDF sets into account. The remainder is estimated by 
varying the $Q_T$ 
distribution of {\sc ResBos} to match the predictions of {\sc pythia}.

$P_u$ and $P_d$ are determined by 
requiring the best agreement between the observed $A_{FB}$ distributions in both the 
dielectron and dimuon events and their corresponding MC 
templates. Since $P_u$ and $P_d$ are simultaneously fitted, 
their values and corresponding uncertainties 
are correlated with a correlation coefficient $\rho = -0.859$. The central value of $R$ and 
its uncertainty are calculated 
using the measured values and the total uncertainties of $P_u$ and $P_d$, and their correlation. 

The measured     
$P_u$, $P_d$ and the ratio $R$ in the full range $|Y|=[0, 2.3]$ is:

\begin{footnotesize}
\begin{eqnarray*}
  P_u &=& 0.602 \pm 0.019 (\text{stat.}) \pm 0.010 (\text{theory}) \pm  0.006 (\text{syst.}) \nonumber \\
             &=& 0.602 \pm 0.022 \nonumber \\
  P_d &=& 0.258 \pm 0.023 (\text{stat.}) \pm 0.012 (\text{theory}) \pm 0.005 (\text{syst.}) \nonumber \\
             &=& 0.258 \pm 0.026 \nonumber \\
  R &=& 2.34 \pm 0.32.
\end{eqnarray*}
\end{footnotesize}

\noindent The systematic uncertainty corresponds to the quadratic sum of the 
uncertainties of imperfect 
efficiency determination, lepton calibration and background estimation. The theory uncertainty 
is the quadratic sum of the uncertainties due to $\Delta$ parameters, QCD calculation and 
fixed value of $\effstw$.  
The systematic and theoretical uncertainties are small compared with the statistical uncertainties. 
Compared with the predictions of CT18NNLO, MSHT20~\cite{MSHT20} and NNPDF4.0~\cite{NNPDF4} 
shown in Table~\ref{tab:average}, the measured $P_u$ is lower than the PDF predictions, while 
$P_d$ is higher. This tendency is consistent with the previous measurement using 
5 fb$^{-1}$ of D0 data~\cite{CMSD0Extraction}, where the mass distribution was 
unfolded. 
In the current analysis, the $P_u$ and $P_d$ parameters are measured by 
comparing the data and the simulated MC, and hence 
there is no unfolding-related uncertainties.
The ratio $R$ is 
lower than the predictions by about 2 standard deviations (the largest difference, 2.8 standard deviations, is 
observed with respect to NNPDF4.0).

\begin{table}[hbt]
\begin{footnotesize}
\begin{tabular}{|l|c|c|c|}
\hline 
     & $P_u$  & $P_d$ & $R$ \\
\hline
 Measured & 0.602$\pm$0.022 & 0.258$\pm$0.026 & 2.34$\pm$0.32 \\
\hline
CT18NNLO & 0.636$\pm$0.011 & 0.213$\pm$0.009 & 2.99$\pm$0.16 \\
\hline
MSHT20 & 0.633$\pm$0.009 & 0.204$\pm$0.008 & 3.10$\pm$0.14 \\
\hline
NNPDF4.0 & 0.624$\pm$0.008 & 0.190$\pm$0.007 & 3.29$\pm$0.13 \\
\hline
\end{tabular}
\caption{\footnotesize Measured values of $P_u$, $P_d$ and $R$ in the full $|Y|$ range $[0, 2.3]$, 
together with their predictions from the CT18NNLO, MSHT20 and NNPDF4.0 PDFs. Predictions are 
calculated using {\sc ResBos} based on the definition in Eq.~\eqref{eq:AFBfactorization} and Eq.~\eqref{eq:averagedef}.
The measured values are presented with their total uncertainties. 
The theoretical predictions are calculated in the same $|Y|$ range and shown with their PDF uncertainties.}
\label{tab:average}
\end{footnotesize}
\end{table} 

\begin{figure}[!hbt]
\begin{center}
\epsfig{scale=0.4, file=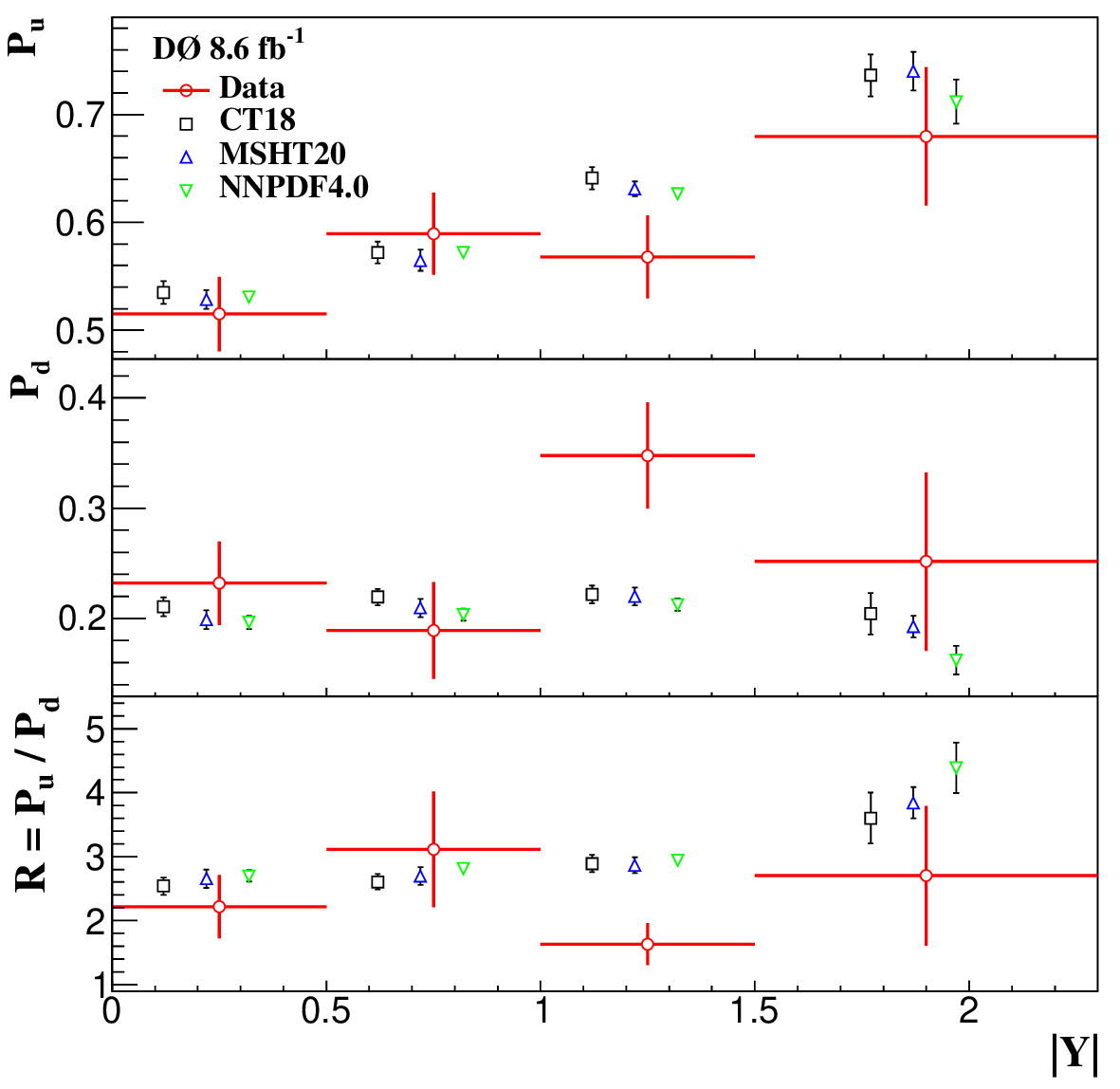}
\caption{\footnotesize Measured values of $P_u$, $P_d$ and $R$ parameters compared with 
the predictions of CT18NNLO, MSHT20 and NNPDF4.0. Error bars of the data points correspond to the total uncertainty of the measurement, while error bars on the predictions correspond to the PDF uncertainties. 
The PDF predictions are offset from the centers of the intervals for clarity.}
\label{fig:ZY}
\end{center}
\end{figure}

The $|Y|$-dependent measurements using both the dielectron and dimuon $A_{FB}$ distributions 
are shown in Table~\ref{tab:ZYresults}. 
The correlation 
coefficients of $P_u$ and $P_d$ in the four $|Y|$ intervals are $-$0.855, $-$0.862, $-$0.866 and $-$0.871 respectively.
The comparison between the measured values and the predictions from representative PDFs is shown in Fig.~\ref{fig:ZY}. 
For $1<|Y|<1.5$ corresponding to $x\sim 0.2$, which is around the peak of the 
parton density distributions of the $u$ and $d$ quarks, 
the measured $R$ differs from the PDF predictions by more than 3.5 standard deviations, 
indicating 
that the $d$ quark contribution is higher than the PDF expectations. 
For the other three bins, the measurements of $P_u$ and $P_d$ show good agreement with the 
predictions.

\begin{table}[hbt]
\begin{footnotesize}
\begin{tabular}{|c|c|c|}
\hline
$|Y|$ range & $P_u$ & $\delta P_u$ \\\hline
 $[0, 0.5]$  & $0.515 \pm 0.031 \pm 0.011 \pm 0.009 \pm 0.004 \pm 0.005$ & 0.034~ \\ 
 $[0.5, 1.0]$ & $0.589 \pm 0.035 \pm 0.010 \pm 0.008 \pm 0.004 \pm 0.005$ & 0.038~ \\ 
 $[1.0, 1.5]$   & $0.568 \pm 0.036 \pm 0.007 \pm 0.010 \pm 0.005 \pm 0.003$ & 0.038~ \\ 
 $[1.5, 2.3]$   & $0.680 \pm 0.060 \pm 0.009 \pm 0.020 \pm 0.005 \pm 0.003$ & 0.064~ \\ \hline \hline
$|Y|$ range & $P_d$ & $\delta P_d$ \\\hline
 $[0, 0.5]$   & $0.232 \pm 0.036 \pm 0.007 \pm 0.007 \pm 0.008 \pm 0.001$  & 0.038 \\ 
 $[0.5, 1.0]$   & $0.189 \pm 0.042 \pm 0.008 \pm 0.007 \pm 0.008 \pm 0.004$  & 0.044 \\ 
 $[1.0, 1.5]$   & $0.348 \pm 0.046 \pm 0.005 \pm 0.008 \pm 0.010 \pm 0.002$  & 0.048 \\  
 $[1.5, 2.3]$   & $0.252 \pm 0.076 \pm 0.014 \pm 0.020 \pm 0.009 \pm 0.002$  & 0.081 \\  \hline \hline
$|Y|$ range & $R$ & $\delta R$\\\hline
 $[0, 0.5]$   & 2.22  & 0.50 \\ 
 $[0.5, 1.0]$   & 3.11  & 0.90 \\ 
 $[1.0, 1.5]$   & 1.63  & 0.33 \\  
 $[1.5, 2.3]$   & 2.70  & 1.09 \\  
\hline 
\end{tabular}
\caption{\footnotesize Measurements of $P_u$, $P_d$ and $R$ in different $|Y|$ bins. 
The uncertainties, in order, are statistical, experimental systematic, 
$\Delta$-induced, $\effstw$ and QCD modelling. The final column is the total uncertainty.}
\label{tab:ZYresults}
\end{footnotesize}
\end{table} 
~\\

The measurements of $P_u$ and $P_d$ for dielectron and dimuon channels 
separately are given in Table~\ref{tab:ZY_eemumu}.
Due to the 
limited detector acceptance and efficiencies for the muons, 
the dimuon events contribute  appreciably only to the 
two lower $|Y|$ intervals. 
For both the $P_u$ and $P_d$ parameters in the two lower $|Y|$ bins, 
the electron and muon measurements agree within 1.7 standard deviations. 

\begin{table}[hbt]
\begin{footnotesize}
\begin{center}
\begin{tabular}{|l|c|c|c|}
\hline
  $|Y|$ range    & & $P_u$  & $\delta P_u$  \\
\hline
\multirow{3}{*}{[0, 0.5]}& $ee$ & $0.554 \pm 0.048 \pm 0.008 \pm 0.010$ & 0.049  \\
     & $\mu\mu$ &  $0.504 \pm 0.041 \pm 0.017 \pm 0.014$ & 0.047 \\
     & CT18NNLO & $0.535\pm0.010$ & \\
          \hline
\multirow{3}{*}{[0.5, 1]} & $ee$ & $0.528 \pm 0.049 \pm 0.010 \pm 0.010$ & 0.051 \\
           & $\mu\mu$ & $0.656 \pm 0.054 \pm 0.017 \pm 0.013$ & 0.058 \\
           & CT18NNLO & $0.572\pm0.010$ & \\
\hline
\hline
  $|Y|$ range    & & $P_d$  & $\delta P_d$ \\
\hline
\multirow{3}{*}{[0, 0.5]} & $ee$ & $0.143 \pm 0.063 \pm 0.004 \pm 0.010$ & 0.064  \\
                & $\mu\mu$ & $0.266 \pm 0.044 \pm 0.012 \pm 0.012$ & 0.047 \\
                & CT18NNLO & $0.211\pm0.008$ & \\
\hline
\multirow{3}{*}{[0.5, 1]} & $ee$ & $0.270 \pm 0.066 \pm 0.007 \pm 0.011$ & 0.067 \\
          & $\mu\mu$ & $0.124 \pm 0.055 \pm 0.013 \pm 0.012$ & 0.058 \\
          & CT18NNLO & $0.220\pm0.007$ & \\
\hline
\end{tabular}
\caption{\footnotesize Central values and uncertainties of the $|Y|$-dependent $P_u$ and $P_d$ parameters 
using dielectron events and dimuon events.
 The uncertainties, in order, are statistical, experimental systematics and theoretical systematics including PDF, $\effstw$ and QCD modelling. 
 The last column $\delta P_q$ gives the total uncertainty.
 Predictions of CT18NNLO are shown with corresponding PDF uncertainties.} 
\label{tab:ZY_eemumu}
\end{center}
\end{footnotesize}
\end{table}

In conclusion, we have performed a new measurement of the proton structure parameters $P_u$ and $P_d$ using 
the forward-backward asymmetry in 
$p\bar{p}\rightarrow Z/\gamma^* \rightarrow \ell^+\ell^-$ events 
using Tevatron data corresponding to 8.6 fb$^{-1}$ of integrated luminosity. 
Taking advantage of the 
asymmetry of the weak interaction, 
the $u$ and $d$ quark contributions are determined separately.
For $p\bar{p}$ collisions at $\sqrt{s}=1.96$ TeV, 
$P_u$ and $P_d$ are dominated by the valence $u$ and $d$ quarks for $0.004<x<0.45$. 
$P_u$, $P_d$ and their ratio $R$ are measured both for the dilepton rapidity 
interval $|Y|=[0, 2.3]$, and 
for finer $|Y|$ intervals 
to investigate their dependence on $x$. 
For the interval $1<|Y|<1.5$, the ratio of $P_u$ and $P_d$ differs from CT18NNLO, 
MSHT20 and NNPDF4.0 PDF predictions by more than 3.5 standard deviations. 
For the other three intervals, 
the results show good agreement with the PDF predictions. 

~\\
This document was prepared by the D0 collaboration using the resources of the Fermi National Accelerator Laboratory (Fermilab),
a U.S. Department of Energy, Office of Science, HEP User Facility. Fermilab is managed by Fermi Research Alliance, LLC (FRA),
acting under Contract No. DE-AC02-07CH11359.

We thank the staffs at Fermilab and collaborating institutions,
and acknowledge support from the
Department of Energy and National Science Foundation (United States of America);
Alternative Energies and Atomic Energy Commission and
National Center for Scientific Research/National Institute of Nuclear and Particle Physics  (France);
Ministry of Education and Science of the Russian Federation, 
National Research Center ``Kurchatov Institute" of the Russian Federation, and 
Russian Foundation for Basic Research  (Russia);
National Council for the Development of Science and Technology and
Carlos Chagas Filho Foundation for the Support of Research in the State of Rio de Janeiro (Brazil);
Department of Atomic Energy and Department of Science and Technology (India);
Administrative Department of Science, Technology and Innovation (Colombia);
National Council of Science and Technology (Mexico);
National Research Foundation of Korea (Korea);
Foundation for Fundamental Research on Matter (The Netherlands);
Science and Technology Facilities Council and The Royal Society (United Kingdom);
Ministry of Education, Youth and Sports (Czech Republic);
Bundesministerium f\"{u}r Bildung und Forschung (Federal Ministry of Education and Research) and 
Deutsche Forschungsgemeinschaft (German Research Foundation) (Germany);
Science Foundation Ireland (Ireland);
Swedish Research Council (Sweden);
and
China Academy of Sciences and National Natural Science Foundation of China (China);
~\\


\begin{thebibliography}{99}
\bibitem{D0eta}
   We use a cylindrical coordinate system with the $z$ axis along the proton beam direction. 
   Pseudo-rapidity is defined as $\eta = -\ln [\tan(\theta_\text{pol}/2)]$ where the polar angle $\theta_\text{pol}$ is 
   measured with respect to the interaction vertex. In the massless limit, $\eta$ is equivalent 
   to the rapidity $y=(1/2) \ln [(E+p_z)/(E-p_z)]$, and $\eta_\text{det}$ is the pseudo-rapidity 
   measured with respect to the center of the detector.
\bibitem{Binning}
   We use a bin size of 1 GeV for the $M$-dependence and a bin size of 0.1 for the $Y$-dependence. For the $Q_T$-dependence, the bin size is 1 GeV, 10 GeV and 100 GeV for $Q_T<10$, $10<Q_T<50$ and $50<Q_T<250$ GeV respectively.
\end{thebibliography}
\end{document}